\documentstyle[11pt,newpasp,twoside,epsf]{article}
\def\edcomment#1{\iffalse\marginpar{\raggedright\sl#1\/}\else\relax\fi}
\reversemarginpar

\begin{document}
\title{Circular Polarization of Circumstellar H$_2$O Masers: Magnetic Fields of AGB stars}
 \author{Wouter Vlemmings}
\affil{Cornell University, 524 Space Sciences Building, Ithaca, NY 14853-6801, USA}
\author{Phil Diamond}
\affil{Jodrell bank Observatory, University of Manchester,
 Macclesfield, Cheshire, SK11 9DL, England}
\author{Huib Jan van Langevelde}
\affil{Joint Institute for VLBI in Europe, Postbus 2, 7990 AA,
 Dwingeloo, The Netherlands}

\begin{abstract}
We present circular polarization measurements of circumstellar H$_2$O
masers around 2 Mira variables and 4 supergaint stars. Both an LTE and
non-LTE method are used to analyze the circular polarization and total
intensity spectra and to obtain the magnetic field strength. The
non-LTE method is found to be able to reproduce the observations
best. Using this, we find fields from a few hundred milliGauss up to a
few Gauss, indicating a solar-type $r^{-2}$ dependence of the magnetic
field on the distance, which leads to stellar surface magnetic fields
of up to several hundred Gauss. No linear polarization is detected to
less than 1.5\%.
\end{abstract}

\section{Introduction}
\label{intro} 
The role of magnetic fields in the late stages of stellar evolution is
still unclear. Blackman et al. (2001) have shown that AGB stars could
produce fields of several hundreds of Gauss. Such strong fields can
play an important role in driving stellar winds and shaping the
outflows, giving rise to the non-spherical shapes observed in
Planetary Nebulae.

 Until recently, information on the magnetic field in the
circumstellar envelopes was obtained by polarimetric observations of
SiO masers at $\approx~2-4~R_*$ from the central star, and OH maser at
a distance of $1000 - 10000$~AU. The SiO observations indicated fields
of 5-10~Gauss for Mira stars and up to 100~Gauss for supergiants
(e.g. Barvainis et al., 1987). However, using a non-Zeeman
interpretation of the observed circular polarization, the magnetic
fields could be a factor 1000 less (Wiebe \& Watson, 1998). The OH
maser observations indicated field strengths of 1-2~mG (e.g. Szymczak
\& Cohen, 1997).

 Now we have determined the magnetic field strengths in the H$_2$O
maser region, at a few hundred AU. Although H$_2$O is a
non-paramagnetic molecule, it has been possible to observe the
circular polarization on some of the strongest circumstellar H$_2$O
maser features. We have used both the LTE analysis presented in
Vlemmings et al. (2001) and non-LTE models based on the models
presented in Neduloha \& Watson (1990). A full description of the
analysis methods, observations and results are presented in Vlemmings
et al. (2002).

\section{Observations}
\label{obs}

We have observed 8 late-type stars with the VLBA. The first
observations, on December 13th 1998, were performed on the supergiants
S~Per, VY~CMa and NML~Cyg, and the Mira variable star U~Her. Recently,
on May 20th 2003, additional observations were performed on the Mira
variable stars U~Her, U~Ori and R~Cas, and the supergiant VX~Sgr. To
get the highest spectral resolution, required for the circular
polarization measurements, the data were correlated twice. Once with
modest spectral resolution ($0.1$ km/s), to get all 4 polarization
combinations (RR, LL, RL and LR), and once with high resolution
($0.027$ km/s), with only RR and LL. The calibration was mainly
performed on the modest spectral resolution data and the solutions
were copied and applied to the high resolution data. This data set was
then used to produce circular polarization and total intensity image
cubes. The modest resolution data set was used to determine the linear
polarization.

\begin{figure} 
   \plotone{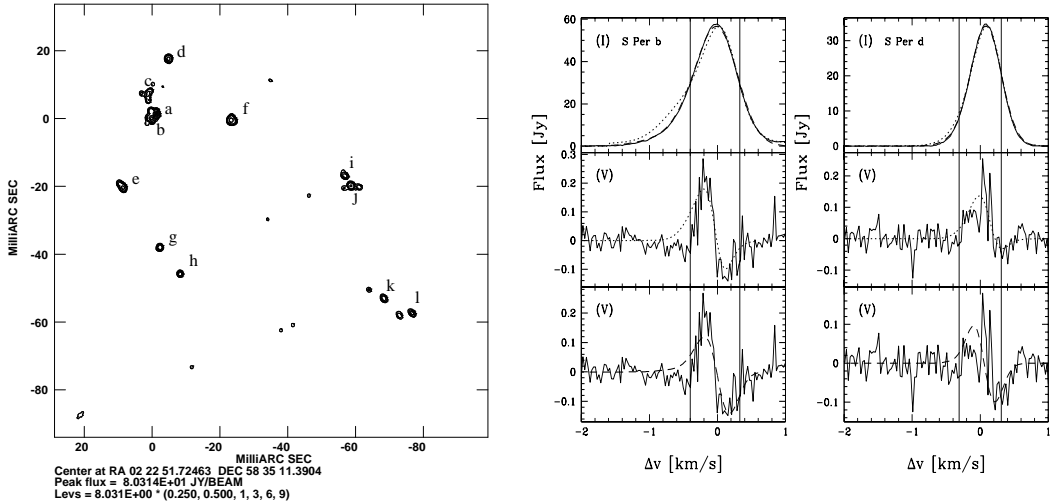}
   \caption{(left) Total intensity image of the H$_2$O maser features
  around S~Per. (right) Total power (I) and V-spectra for selected
  maser features of S~Per. The bottom panel shows the best fitting
  LTE model (dashed), the middle panel shows the best fitting non-LTE
  model (dotted).}
   \label{sper}
\end{figure}

\section{Results}
\label{res}

 We have examined the strongest H$_2$O maser features around the 7
stars observed. Circular polarization between $0.1\%$ and $25\%$ of
the total intensity was detected on $\approx 50\%$ of the brightest
maser features. An example of the features around S~Per is shown in
Fig.1.  We rule out any systematic effects as a cause of the observed
spectrum, because various percentages of circular polarization are
observed as well as different directions of the magnetic field.  No
linear polarization was detected in excess of $\approx 1.5\%$.

 The magnetic field strengths were determined with both the LTE and
the non-LTE method. The LTE method predicts the circular polarization
spectrum to be directly proportional to the derivative of the total
power spectrum. We found that the observed spectra were narrower,
which can only be explained with the non-LTE analysis. The non-LTE
field strengths are $\approx 40\%$ lower than those determined by the
LTE method.

From the observations and analysis, we estimate the magnetic field
strengths in the H$_2$O maser region to be $\approx 200$~mG for S~Per
and VY~CMa. The field around NML~Cyg is $\approx 500$~mG, while the
Mira variable U~Her shows a much higher field of $\approx 1.5$~G.
Preliminary results for the second set of observations indicate a
magnetic field of up to several Gauss on VX~Sgr and
U~Ori. Unfortunately the masers around R~Cas were not detected. The
second observations of U~Her provide an upper limit to the magnetic
field of several hundred mG on the observed features. Since the
velocity of the features observed around U~Her at the observation dates are
significantly different, we are now observing different maser spots.
These spots are likely further out in the circumstellar envelope and
have larger angles between line of sight and the magnetic field, which
results in a lower observed field strength.

\section{Discussion}
\label{disc}

 Our results favor the non-LTE approximation and because we do not
detect any linear polarization, a non-Zeeman interpretation is also
highly unlikely. The lack of linear polarization can be easily
explained in the non-LTE case, because linear polarization is only
produced by strongly saturated masers. A line widths analysis
indicates that the circumstellar H$_2$O masers are not saturated. Even
for large angles between the line of sight along the maser and the
direction of the magnetic field we do not expect any linear
polarization. In the LTE analysis, the lack of linear polarization can
only be explained by having the maser line of sight beam along the
magnetic field lines.

 We can compare the strength of the magnetic field in the H$_2$O maser
region with the values obtained from SiO and OH maser polarization
observations. This seems to indicate that the magnetic field strength
values inferred from the SiO maser observations are indeed due to the
normal Zeeman effect, although Elitzur (1996) has argued that the
field strength can still be a factor 10 lower on both SiO and OH
masers. Fig.2 shows the dependence of the magnetic field strength on
distance from the star. Our observed values are plotted at the
observed maximum extent of the H$_2$O maser region. Because our
observations are most sensitive for the highest magnetic fields, we
are actually probing the inner edges of the H$_2$O maser shell. The
arrows indicate the typical thickness of such a shell. These results
indicates that the magnetic field strength is best represented by a
solar-type dependence on distance ($r^{-2}$). The exact shape of the
magnetic field strength cannot easily be determined from our
observations. The SiO polarization maps indicate mostly radial field
lines close to the star.

\begin{figure}[t!] 
%   \plotfiddle{vlemmings_fig2.ps}{7truecm}{0}{40}{40}{-140}{-60}
  \plotone{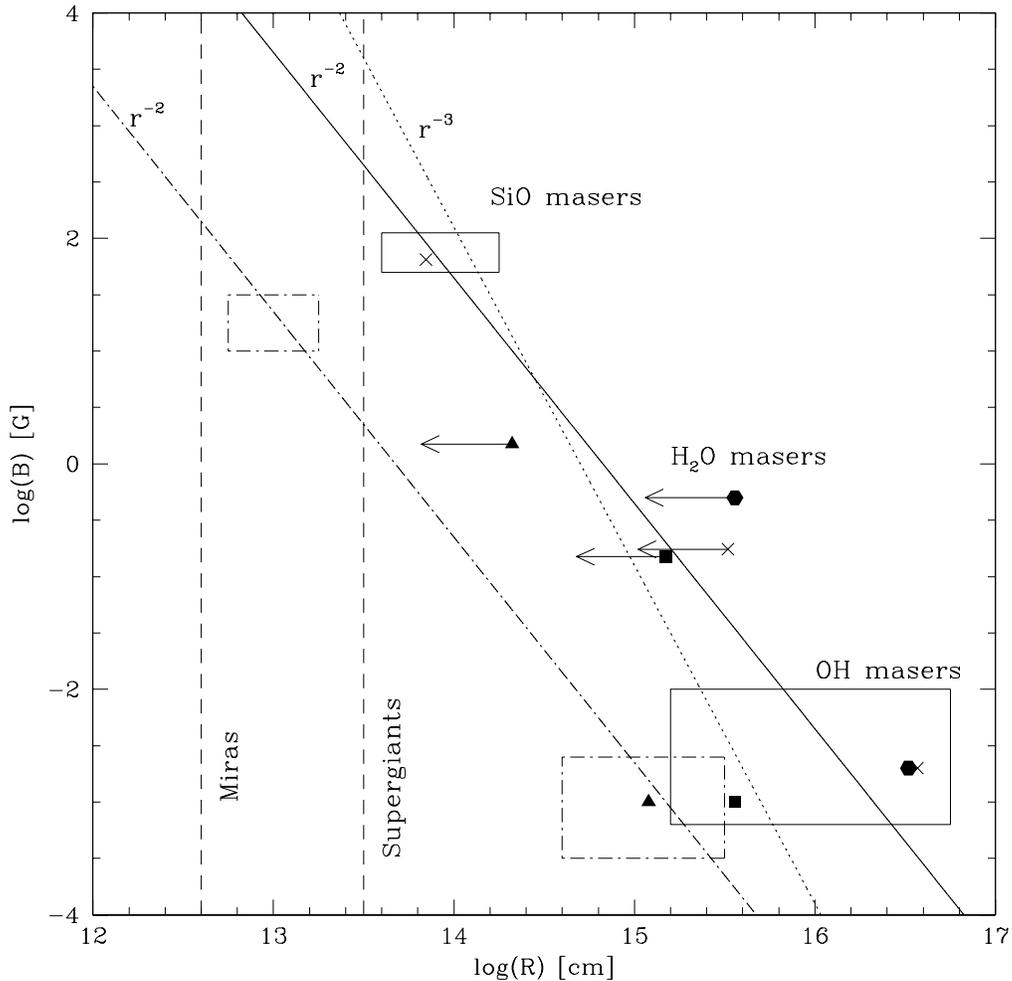}
    \caption{Magnetic field strength B, as function of distance R from
the star. Dashed-dotted boxes are the SiO and OH maser estimates for
Mira stars, solid boxes are those for supergiant stars. Symbols
indicate observations (Dec 1998); U~Her is represented by triangles,
S~Per by the square, VY~CMa by the crosses and NML~Cyg by the
hexagonal symbol. The dashed vertical lines are an estimate of the
stellar radius.}
   \label{rb}
\end{figure}

The magnetic pressure of the field in the H$_2$O maser region
dominates the thermal pressure by a factor of 20. Using the solar-type
field, extrapolated surface field strengths are of the order of
$100-1000$ Gauss, strong enough to drive and shape the outflows.

\end{document}